\documentclass[sigconf]{acmart} 
\AtBeginDocument{%
  }

\setcopyright{acmlicensed}
\usepackage{enumitem}

\copyrightyear{2018}
\acmYear{2018}
\acmDOI{XXXXXXX.XXXXXXX}
\acmConference[Conference acronym 'XX]{Make sure to enter the correct
  conference title from your rights confirmation email}{June 03--05,
  2018}{Woodstock, NY}
\acmISBN{978-1-4503-XXXX-X/2018/06}

\begin{document}

\title{MusicScaffold: Bridging Machine Efficiency and Human Growth in Adolescent Creative Education through Generative AI}

\author{Zhejing Hu}
\email{zhejing.hu@connect.polyu.hk}
\affiliation{%
  \institution{The Hong Kong Polytechnic University}
  \city{Hong Kong}
  \country{China}
}

\author{Yan Liu}
\email{yan.liu@polyu.edu.hk}
\affiliation{%
  \institution{The Hong Kong Polytechnic University}
  \city{Hong Kong}
  \country{China}
}

\author{Zhi Zhang}
\email{zhi271.zhang@connect.polyu.hk}
\affiliation{%
  \institution{The Hong Kong Polytechnic University}
  \city{Hong Kong}
  \country{China}
}

\author{Gong Chen}
\email{heinz@clozzz.com}
\affiliation{%
  \institution{FireTorch Partners}
  \city{Hong Kong}
  \country{China}
}

\author{Bruce X.B. Yu}
\email{xinboyu@intl.zju.edu.cn}
\affiliation{%
  \institution{ZJU-UIUC Institute, Zhejiang University}
  \state{Zhejiang}
  \country{China}
}
\author{Junxian Li}
\email{junxianli@gmail.com}
\affiliation{%
  \institution{Shenzhen International Foundation College}
  \state{Shenzhen}
  \country{China}
}

\author{Jiannong Cao}
\email{csjcao@comp.polyu.edu.hk}
\affiliation{%
  \institution{The Hong Kong Polytechnic University}
  \city{Hong Kong}
  \country{China}
}

\renewcommand{\shortauthors}{Trovato et al.}

\begin{abstract}
Adolescence is marked by strong creative impulses but limited strategies for structured expression, often leading to frustration or disengagement. While generative AI lowers technical barriers and delivers efficient outputs, its role in fostering adolescents’ expressive growth has been overlooked. We propose \emph{MusicScaffold}, the first adolescent-centered framework that repositions AI as a guide, coach, and partner, making expressive strategies transparent and learnable, and supporting autonomy. In a four-week study with middle school students (ages 12–14), MusicScaffold enhanced cognitive specificity, behavioral self-regulation, and affective confidence in music creation. By reframing generative AI as a scaffold rather than a generator, this work bridges the machine efficiency of generative systems with human growth in adolescent creative education.
\end{abstract}




\received{20 February 2007}
\received[revised]{12 March 2009}
\received[accepted]{5 June 2009}

\begin{teaserfigure}
  \centering
  \includegraphics[width=\linewidth,trim=0 0 0 0,clip]{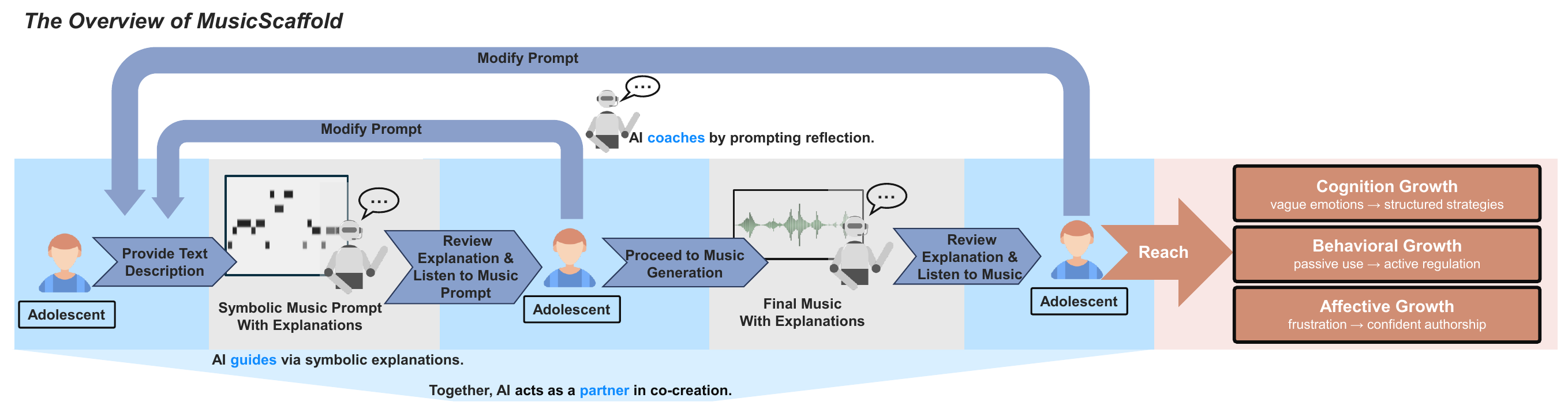}
  \Description{This figure presents the conceptual structure of MusicScaffold. 
        On the left side, adolescent learners provide vague natural-language descriptions of musical intentions. 
        On the right side, the framework processes these inputs through three complementary AI roles: 
        (1) as a guide, it translates vague emotional words into symbolic attributes such as key, tempo, or timbre with plain-language explanations; 
        (2) as a coach, it supports iterative refinement by prompting learners to reflect on why specific structural changes affect expression; 
        (3) as a partner, it generates full musical outputs while ensuring that learners retain authorship and creative control. 
        At the bottom, the framework is aligned with three evaluation dimensions: 
        RQ1 (Cognitive), which examines whether students can move from vague to structured expression; 
        RQ2 (Behavioral), which investigates whether they shift from passive reliance on outputs to active regulation of strategies; 
        and RQ3 (Affective), which explores whether they develop autonomy, intrinsic motivation, and confidence in creative expression.}
  \caption{Overview of MusicScaffold. The framework positions generative AI as a hybrid learning companion with three complementary roles. A guide that provides symbolic explanations linking emotions to musical structures, a coach that prompts reflection during iterative refinement, and a partner that supports co-creation while safeguarding learner autonomy. These roles were evaluated across three dimensions of adolescent engagement: RQ1 (Cognitive), RQ2 (Behavioral), and RQ3 (Affective).}
  \label{fig:intro}
\end{teaserfigure}
\maketitle
\section{Introduction}

Adolescence is often characterized as a turbulent and rebellious stage of development \cite{hall1905adolescence,chowdhury2023co}. During this period, emotional experiences are both intense and abundant, and adolescents often display a strong desire for creativity and self-expression \cite{egana2023impacts,ishiguro2023extracurricular,long2024parallel,long2024xylocode}. Yet their expressive methods and strategies remain immature in that they long to convey complex emotions but often lack the knowledge to transform vague impulses into structured forms of expression \cite{stevenson2014training,celume2019fostering,zielinska2022adolescents,murgia2023chatgpt}. This gap, in which adolescents want to express but do not know how, can lead to frustration and even disengagement from artistic activities, such as abandoning a task after being unable to match a vague emotional idea with a concrete expressive strategy \cite{ryan2000self,rostan2010studio,anderson2020mistakes}. Conversely, when effective supports for expression are provided, adolescents' creativity is amplified rather than suppressed \cite{deci2012self,beghetto2014classroom}. The development of structured expressions is thus essential, as it enables raw emotional impulses to be transformed into concrete creative outcomes and marks a developmental leap from spontaneous inspiration to controllable, deliberate creativity \cite{amabile2018creativity,mumford2013creative,beghetto2014classroom}.

Generative AI dramatically reduces the technical barriers to artistic expression, enabling adolescents to create across text \cite{qin2024charactermeet,chakrabarty2024art,goldi2024intelligent,weber2024legalwriter,ali2025even}, visual art \cite{chen2024videocrafter2,wang2024promptcharm,mahdavi2024ai}, and music \cite{choi2025exploring,kamath2024sound,vear2024jess+} from simple descriptions.  Through instant feedback, these systems foster vivid emotional externalization and sustained practice without prolonged training \cite{newman2024want,han2024teachers,frich2019mapping,lin2020your,muller2022genaichi,chefer2023attend,wadinambiarachchi2024effects,bird2025diffraction,lu2025project}. In practice, tools such as Suno \cite{suno2024} and Udio \cite{udio} (music), GPT series \cite{gpt4o} (writing), and MidJourney \cite{midjourney} (visual art) exemplify this potential by enabling fluent outputs or iterative refinements, while mixed-initiative approaches add explanatory guidance \cite{amershi2019guidelines,rezwana2023designing,chen2024investigating,krol2025exploring}. Together, these advances demonstrate the value of generative AI in improving accessibility, engagement, and the educational relevance of creative practices. In music, emotional intentions map directly onto formal structures such as key, tempo, and harmony \cite{choi2025understanding,cavez2025euterpen,cavez2024challenges,lee2025mvprompt,morrison2024entangling}, which makes it a natural testbed for developing scaffolds that render expressive strategies transparent and learnable. Yet this pedagogical challenge remains open, particularly as efficiency-driven use can foster dependency and reduce initiative \cite{ahmad2023impact}.

Building on these advances, we propose \emph{MusicScaffold} (Fig. \ref{intro}), the first adolescent-centered interaction framework positioned as a hybrid learning companion. It serves simultaneously as a \emph{guide} by providing symbolic explanations, a \emph{coach} by prompting reflection during iterative refinement, and a \emph{partner} by supporting co-creation while safeguarding learner autonomy. MusicScaffold establishes a closed-loop of support through two mechanisms:
(1) Symbolic prompt explanation (guide role) decomposes vague emotional intentions (e.g., ``happiness'') into adjustable expressive elements (e.g., major key, fast tempo, piano timbre), helping learners connect emotions to structure and overcome the challenge of ``not knowing how to translate impulses''.
(2) Iterative human–AI reflection (coach role) encourages learners to examine why particular structures convey their intent, shifting them from ``passively accepting outputs'' to ``actively understanding expressive logic'' and thereby cultivating strategic refinement.
Together, these mechanisms are grounded in the overarching role of a partner, ensuring that adolescents retain creative direction while avoiding the loss of autonomy that can result from excessive intervention.

We evaluated MusicScaffold through a mixed-methods study that integrated multi-round creative tasks, interaction logs, questionnaires, and interviews. Our analysis focused on three research questions (RQ), organized around the cognitive, behavioral, and affective dimensions of student engagement \cite{fredricks2004school,sinatra2015challenges,wei2021assessment}.
\begin{itemize}[left=0pt]
    \item Cognitive dimension (RQ1): Can structured symbolic prompts help adolescents move from vague, single-dimensional expressions to precise, multi-dimensional outcomes, thereby balancing immediate creative efficiency with long-term cognitive growth? 
    \item Behavioral dimension (RQ2): Can symbolic prompt representations enable adolescents to regulate and refine their creative process—shifting from passive reliance on AI to active strategy building?
    \item Affective dimension (RQ3): Can interactive guidance strengthen adolescents’ autonomy and intrinsic motivation, turning efficient AI outputs into opportunities for sustained creative growth?
\end{itemize}

The results show that MusicScaffold not only helps adolescents refine vague ideas into structured outcomes but also shifts their behavior from passive reception to active regulation. Its core value lies in positioning AI as a hybrid learning companion, creating synergy between AI’s capacity for instant generation and adolescents’ long-term expressive growth. This exploration points to a new direction for applying generative AI in adolescent creativity support, moving from empowering creation to empowering growth.

Our paper makes three key contributions:
\begin{itemize}[left=0pt]
    \item We reframe adolescents’ strong creative impulses alongside their limited strategies for structured expression as a critical tension between generative AI’s efficiency and learners’ growth. This perspective advances HCI’s understanding of how creativity support tools can balance immediate accessibility with long-term skill development.
    \item We propose MusicScaffold, the first adolescent-centered framework that positions AI as a guide, coach, and partner to scaffold expressive strategies rather than only generate outputs. This design introduces a new interaction paradigm where generative AI acts as a learning companion rather than a one-shot generator.
    \item We conducted mixed-methods studies with 330 students, including an exploratory study and a comparative study, and the results show that MusicScaffold enhances adolescents’ cognitive specificity, behavioral regulation, and affective autonomy in creative expression. These findings provide evidence that scaffolding-based AI design can cultivate both expressive competence and intrinsic motivation.
\end{itemize}

\section{Related Work}
\subsection{Scaffolding and Adolescent Cognitive Development}
Classic work in educational psychology emphasizes that adolescents learn most effectively when provided with structured support that bridges the gap between intuitive ideas and formal strategies. Vygotsky’s notion of the zone of proximal development and the concept of scaffolding \cite{vygotsky1978mind,wood1976role,bruner1974toward} highlight how learners progress when external guidance makes implicit reasoning explicit. During adolescence, the development of metacognition, domain knowledge, and strategic thinking is still emerging; students often possess strong creative impulses but lack the tools to transform vague feelings into structured expression \cite{perez2019self, zimmerman2002becoming,amabile2018creativity}. Without adequate guidance, this gap can lead to frustration, superficial outcomes, or disengagement. Conversely, when adolescents receive scaffolds that externalize reasoning—such as explanations, structured choices, or reflective prompts—they are more likely to acquire expressive competence and autonomy. Technological systems illustrate this principle. For example, intelligent tutoring systems support learning not by providing solutions directly, but by guiding students through planning, monitoring, and reflection \cite{shneiderman2007creativity,amershi2019guidelines}. These examples show why adolescents need not just access to creative tools but structured pathways that help them connect intentions with strategies. 

While adolescents have been widely studied in HCI across diverse domains \cite{lee2025understanding,wu2024designing,xu2024obviously,kitson2024call,hartwig2024adolescents,liu2024wrist,smout2023enabling}, little attention has been paid to how scaffolding can support their creative expression. Our work addresses this gap by leveraging scaffolding theory to design support that helps adolescents transform vague impulses into structured expressive strategies.

\subsection{Generative AI for Adolescent Creativity}
Generative AI has recently lowered the barrier for young people to create text, images, and music. Interfaces such as sketch-based or conversational prompts \cite{fan2024storyprompt,zhang2022storydrawer,ye2025colin,newman2024want, hedderich2024piece} and classroom platforms that teach AI literacy \cite{dennison2024consumers,zhang2024effectiveness,li2025unseen} show how accessible modalities can spark engagement and self-efficacy. Yet empirical studies also warn that students may remain passive consumers of outputs, without internalizing the reasoning or strategies behind them \cite{denny2024generative}. This limitation is particularly salient in music, where adolescents can generate full songs instantly but struggle to connect fuzzy intentions with structured constructs such as key, tempo, or harmony \cite{choi2025understanding,choi2025exploring,choi2024way}. 

Prototypes boost short-term engagement but rarely measure whether learners gain transferable knowledge or strategic awareness \cite{cai2025child}. In other words, current systems succeed at ``letting adolescents create'' but not at ``helping them learn to express''. In particular, scaffolds that make expressive strategies transparent and learnable remain underexplored, despite evidence that efficiency-driven AI use can sometimes foster dependency and reduce initiative in student populations \cite{ahmad2023impact}. MusicScaffold directly addresses this gap by embedding explanatory and iterative refinement into generative workflows, so that creativity becomes a vehicle for growth rather than just production.

\subsection{Interaction Paradigms and the Case for Music}
Early advances in generative AI demonstrated that a single text prompt could already produce fluent prose, realistic images, or music with minimal effort \cite{brown2020language,ramesh2021zero}. Building on this, subsequent work sought to improve controllability: some introduced interface-level scaffolds that expose latent parameters for parameterized editing \cite{hertz2022prompt,brooks2023instructpix2pix,zhang2023adding,zhuang2023dreameditor,cavez2025euterpen,cavez2024challenges}; others explored dialogic or iterative feedback where users and models co-adjust outputs step by step \cite{wu2023visual,dennison2024consumers,madaan2023self,krol2025exploring}; and more recent studies proposed planning-first pipelines that externalize reasoning before generation \cite{yao2023react,yao2023tree,besta2024graph,xi2023self,lee2025mvprompt,morrison2024entangling}. These approaches show how interaction design can move beyond one-shot generation. Yet the dominant orientation remains output quality, with little attention to whether users, especially adolescents, develop lasting expressive strategies.

Music provides a uniquely suitable testbed for filling this gap. Unlike text or images, music tightly couples emotional intention with structured formal elements (key, rhythm, harmony), offering clear levers for scaffolding strategy-building \cite{bhandari2025text2midi,lu2023musecoco,agostinelli2023musiclm,liu2024audioldm,copet2024simple,wu2024music,melechovsky2023mustango}. Despite rapid advances in text-to-music models (e.g., Suno \cite{suno2024}, Udio \cite{udio}), most systems prioritize fidelity and controllability rather than pedagogy. Our uniqueness lies in reframing music generation as both an expressive outlet and a scaffolded learning environment, where symbolic prompt explanation and iterative refinement help adolescents move from raw feelings to structured, learnable forms of creativity.

\section{Exploratory Study of the Efficiency–Growth Challenge in Prompt-Output AI Music Generation Tools}

\subsection{Motivation}
AI music generation tools such as Suno \cite{suno2024} and Udio \cite{udio} have recently entered creative education, offering adolescents unprecedented efficiency in music generation. Their prompt––output paradigm enables complete pieces with minimal training, lowering barriers to entry and making creativity more immediately accessible. Yet precisely this efficiency raises a central question: \emph{do such tools merely deliver outputs, or do they also foster the growth of expressive competence?} If AI music generation tools bypass the reflective process of translating vague intentions or emotions into structured musical strategies, they risk cultivating dependence on generation rather than developing strategic creativity. This tension between efficiency and growth motivates an exploratory study of how adolescents actually engage with current tools.

To examine this tension, we conducted a four-week study in middle school art classes with 60 students (ages 12–14), where adolescents used commercial AI music generation tools in their regular lessons. We focus on three guiding questions:

\begin{itemize}[left=0pt]
    \item RQ1: When working with current AI music generation tools, can adolescents move beyond vague creative intentions (e.g., emotions, moods) to structured musical elements (e.g., rhythm, mode, harmony), or does efficiency overshadow growth in expressive strategies?
    \item RQ2: How does the direct prompt––output paradigm shape adolescents’ approaches to exploration and refinement—does it encourage deliberate regulation of their creative process, or foster reliance on surface-level regeneration?
    \item RQ3: What effects do current AI music generation tools have on adolescents’ sense of creative autonomy and self-efficacy—do efficient results reinforce their confidence in creative expression, or undermine their belief in their own expressive abilities?
\end{itemize}

\subsection{Study Design}
\subsubsection{Context and Duration}
The exploratory study was conducted over four consecutive weeks in two parallel music classes ($N=60$, ages 12–14) at the same middle school. Both classes were taught by the same music teacher using identical lesson plans, ensuring instructional consistency. Prior to the study, students in both classes completed a short baseline task (describing how they would express ``happiness'' in music) and a brief self-report survey on prior musical training. No significant differences were observed across classes in baseline expressive ability or musical background, supporting their comparability.  

Each week, the teacher guided students through a 30–minute AI-assisted music creation session conducted on electronic tablets, which were standard equipment in daily lessons. To ensure that results reflected the typical prompt-output paradigm rather than idiosyncratic interface features of a single product, we selected two representative commercial platforms—Suno \cite{suno2024} and Udio \cite{udio}. This allowed us to embed both models within a unified classroom interface, ensuring consistent interaction design across groups. Both systems adopt the same paradigm: students type a free-form text prompt and receive a one-shot audio generation, with only superficial controls for re-generation or template switching. Neither provides symbolic prompt explanation of attributes or iterative refinement. Their use therefore captures the current state of accessible prompt-driven GenAI for music creation, while our controlled interface ensured comparability across conditions. This setup also enabled us to examine directly how efficiency in AI-assisted music creation intersects with adolescents’ opportunities for cognitive, behavioral, and affective growth.

In addition to the in–class sessions, the teacher informed students that they could optionally engage in an after–class activity: listening to anonymized peer-generated pieces on the platform and guessing the original input prompts. This activity was voluntary, framed as a supplementary opportunity for reflection and peer learning rather than a graded homework task.

\subsubsection{Weekly Procedure}
Each session consisted of three in–class stages—task introduction, student creation, and feedback collection—followed by an optional after–class activity. Table~\ref{tab:procedure} outlines the sequence and corresponding data sources.

\begin{table*}[h]
\centering
\caption{Weekly classroom procedure and corresponding data collection.}
\begin{tabular}{p{3cm}p{10cm}}
\toprule
\textbf{Stage} & \textbf{Description and Data Collected} \\
\midrule
Task introduction & The teacher presents an open–ended creative theme (e.g., ``Express a rainy morning through music'') and explains that students will use the AI music generation tools to generate music. No data are collected at this stage. \\
\midrule
Student creation & Students independently use AI music generation tools (Suno \cite{suno2024} and Udio \cite{udio}) on their tablets to complete the task. During this stage, all interaction data are logged, including: (1) Prompts: All input text requests; (2) Adjustment behavior: Each revision coded as regeneration, template switching, platform switching, or description modification; (3) Output: Final audio files paired with their originating prompts.   \\
\midrule
Feedback collection & At the end of class, students complete a short survey with scaled and open–ended items (e.g., ``Were you able to express your feelings through music?’’ on a 1–5 scale; ``If you wanted the music to be richer, what would you change?’’ as open response). \\
\midrule
After–class activity & Outside class, students may access an anonymized shared library, listen to peers’ outputs, and submit a guess or comment about the original prompt. This activity was designed as a lightweight social learning opportunity. Participation was voluntary, and interaction logs were collected for analysis. \\
\bottomrule
\end{tabular}
\label{tab:procedure}
\end{table*}

\subsection{Evaluation Dimensions and Metrics}
We operationalized six evaluation dimensions, with each research question linked to two complementary indicators. Together, these measures were designed to capture both the efficiency of producing outputs and the degree of ability growth that occurred through interaction with AI music generation tools.

\begin{figure*}[htbp]
    \centering
    \includegraphics[width=\linewidth,trim=0 0 0 0,clip]{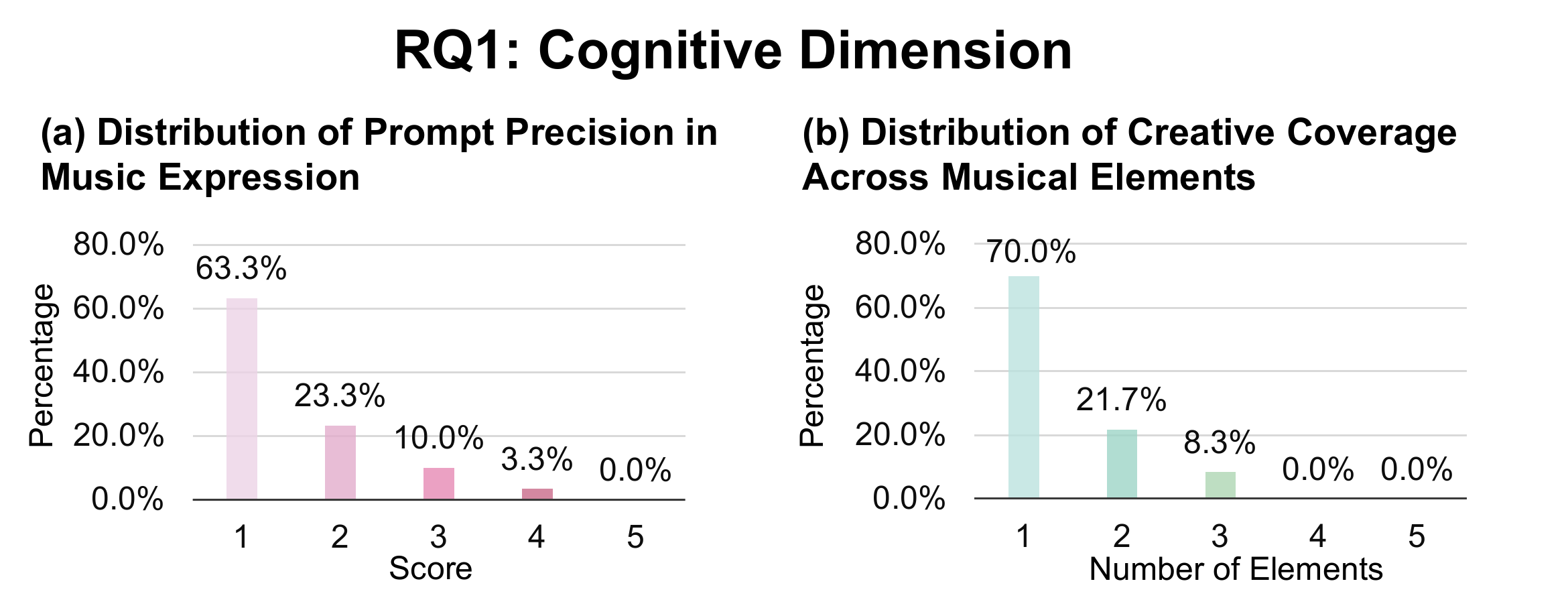}
    \Description{Results for RQ1 showing prompt specificity levels and elemental coverage. Most students stayed at vague, single-dimensional descriptions rather than multi-dimensional structured expressions.}
    \caption{RQ1: Efficiency of outputs overshadows growth in structured expressive strategies. 
    (a) Prompt specificity levels, showing that 86.6\% of prompts were at Levels~1–2 and less than 4\% reached Levels~4–5. 
    (b) Elemental coverage, indicating that most works relied on only one musical dimension.}
    \label{fig:pre1}
\end{figure*}

For RQ1, we examined:

(1) Prompt specificity (Fig.~\ref{fig:pre1}(a)): students’ ability to translate vague ideas into structured musical descriptions. Prompts were rated by the teacher on a 5-point scale, from purely emotional words (e.g., ``happy'', ``sad'') at Level~1, to precise, strategy-driven expressions that explicitly linked musical features with affect (e.g., ``use staccato to show liveliness with added echo effect'') at Level~5.  

(2) Elemental coverage (Fig.~\ref{fig:pre1}(b)): the breadth of musical elements used. Scores ranged from 1 (one element, e.g., rhythm only) to 5 (five or more distinct elements, e.g., rhythm, mode, timbre, melody, and effects combined).

For RQ2, we focused on:  

(3) Adjustment behavior (Fig.~\ref{fig:pre2}(a)): students’ strategies when dissatisfied with generated outputs, categorized into non-strategic adjustments (e.g., direct regeneration, switching templates) and strategic adjustments (e.g., refining prompts, modifying parameters). We calculated the proportion of each adjustment type.  

(4) Prompt homogeneity (Fig.~\ref{fig:pre2}(b)): the teacher reviewed all student prompts (without student identifiers) and rated whether they were homogeneous (i.e., repeated or minimally varied phrasings that closely resembled peers’ prompts) or novel (i.e., distinctive in theme, vocabulary, or expressive intent). We then computed the proportion of homogeneous versus novel prompts.

For RQ3, we measured:  

(5) After-class participation (Fig.~\ref{fig:pre3}(a)): students’ intrinsic motivation beyond classroom tasks, measured as the average number of weekly interactions with peers’ works (e.g., listening and guessing prompts). Mere passive access (e.g., opening the platform without interaction) was not counted.  

(6) Self-efficacy in musical expression (Fig.~\ref{fig:pre3}(b)): students’ perceived confidence in expressing emotions through music, rated on a 5-point scale from 1 (``cannot express at all'') to 5 (``very capable of expressing ideas clearly''). 

\subsection{Results}

\subsubsection{RQ1: Efficiency of outputs overshadows growth in structured expressive strategies}
As shown in Fig.~\ref{fig:pre1} (a), 86.6\% of student prompts fell within Levels 1–2 (pure emotion words or simple modifiers), only 10.0\% reached Level 3 (basic musical terms), and less than 4\% achieved Levels 4–5 (precise, strategy-driven combinations). This indicates that the majority of adolescents still lack the ability to translate vague ideas (e.g., ``happy'') into structured expressions using musical terminology. For example, when attempting to convey ``a feeling of excitement mixed with nervousness'', most students wrote descriptions such as ``a bit excited and scared'', rather than relating it to specific musical strategies like ``dotted rhythm (to enhance movement) and minor modulation (to suggest tension)''.

Fig.~\ref{fig:pre1} (b) further shows that students’ works covered an average of only 1.2 musical dimensions. Seventy percent of compositions focused on a single element such as tempo or volume, while fewer than 9\% incorporated three or more dimensions (e.g., rhythm + mode + timbre). Questionnaire responses aligned with this pattern: 83\% of students admitted they ``wanted the music to be richer but did not know what else to adjust''. One student commented, ``I wanted the melody to feel warmer, but besides slowing down, I didn’t know what to change'', while another noted, ``When I try to express complex feelings, I just keep adjusting the volume, but it never feels right''. These findings confirm that the limitation lies not in the students’ ideas, but in their narrow understanding of how musical elements can function as expressive resources.

\begin{figure*}[htbp]
    \centering
    \includegraphics[width=\linewidth,trim=0 0 0 0,clip]{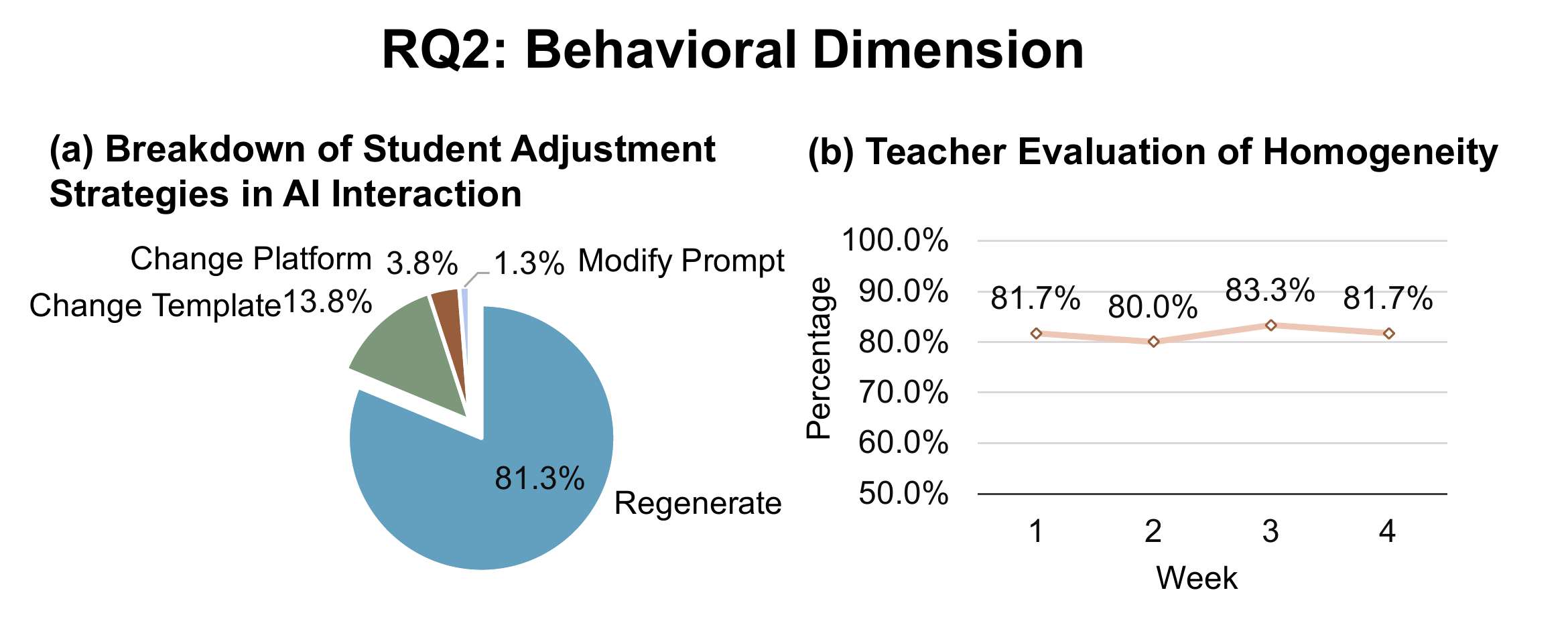}
    \Description{Results for RQ2 showing adjustment behaviors and prompt homogeneity. Students mainly regenerated outputs without modification, and most outputs remained homogeneous across weeks.}
    \caption{RQ2: Reliance on efficient regeneration inhibits deeper strategy development. 
    (a) Adjustment behavior distribution, showing over 80\% of actions were simple regeneration without modification. 
    (b) Prompt homogeneity, where more than 70\% of prompts showed little variation across four weeks.}
    \label{fig:pre2}
\end{figure*}

\subsubsection{RQ2: Reliance on efficient regeneration inhibits the development of deliberate prompt refinements and expressive strategies}
Fig.~\ref{fig:pre2} (a) presents the distribution of adjustment behaviors when students were dissatisfied with generated results. A striking 81.3\% of adjustments were ``regenerate without modification'', while 13.8\% involved simply switching pre-defined templates (e.g., ``happy–pop'' to ``happy–piano''). In contrast, only 1.3\% of behaviors involved strategic adjustments, such as refining prompts (e.g., ``add violin timbre'') or changing parameters (e.g., tempo or key).  

Teacher evaluations in Fig.~\ref{fig:pre2} (b) further revealed that more than 70\% of outputs remained highly homogeneous with respect to initial input descriptions across all four weeks, showing little evidence of creative extension beyond the first idea. Although 60\% of students agreed that AI music generation tools ``helped them complete their work'', follow-up responses suggested a narrow interpretation: e.g., ``They helped me finish the teacher’s assignment'', or ``They generated music that matched my words''. These statements indicate that the perceived benefit was about completing tasks efficiently, rather than developing expressive strategies or deeper creative understanding. In other words, the direct input–output paradigm reinforced reliance on tools, discouraging students from actively exploring how to refine or extend their ideas.

\begin{figure*}[htbp]
    \centering
    \includegraphics[width=\linewidth,trim=0 0 0 0,clip]{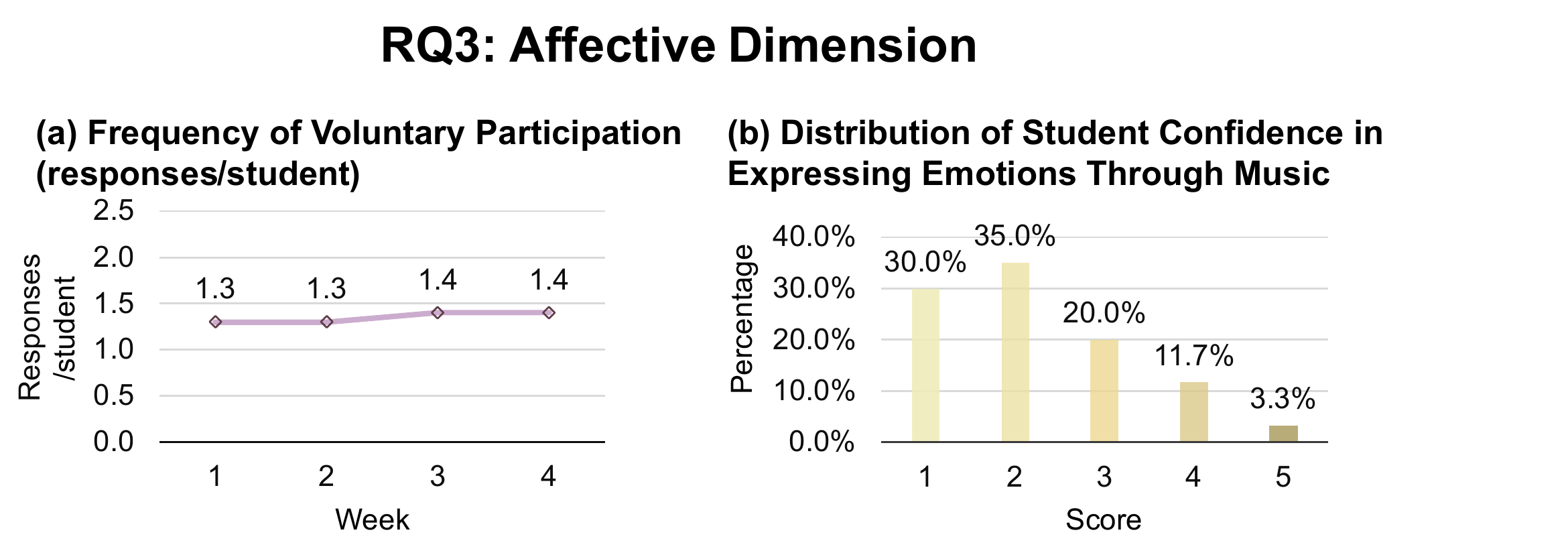}
    \Description{Results for RQ3 showing low after-class participation and weak self-efficacy. Few students voluntarily engaged, and confidence in musical expression remained low.}
    \caption{RQ3: Fast completion undermines autonomy and sustained self-efficacy. 
    (a) After-class participation, with students submitting on average only 1.3 after-class interactions per week. 
    (b) Self-efficacy ratings, where 65\% of students selected 1–2 points, indicating low confidence in their ability to express through music.}
    \label{fig:pre3}
\end{figure*}

\subsubsection{RQ3: Fast completion undermines autonomy and sustained self-efficacy}
The effects of current AI music generation tools on autonomy and self-efficacy appeared limited. As Fig.~\ref{fig:pre3} (a) shows, in a voluntary ``campus life'' themed competition, students submitted on average only 1.3 after-class interactions per week, with no increase across four weeks. This suggests that without external requirements, students showed little intrinsic motivation to use the tools for self-expression. Interviews reinforced this observation, with one student remarking, ``All the outputs sound kind of similar, so I don’t feel like spending extra time on it''.

Fig.~\ref{fig:pre3} (b) presents students’ self-efficacy ratings for ``expressing feelings through music''. Sixty-five percent selected 1–2 points (``not capable'' or ``barely capable''), and overall confidence remained low. Many students attributed expressive success entirely to the tool rather than themselves. As one explained: ``Without the tool, I can’t even think of how to show my ideas in music. Even with the tool, it doesn’t feel like it’s really what I wanted''. Such statements reveal an erosion of belief in their own creative ability.

Taken together, these findings indicate that direct prompt-output AI music generation tools not only fail to improve adolescents’ autonomy and confidence, but may also foster dependency and homogeneous outputs. By reducing opportunities for reflective strategy use, the tools risk reinforcing a cycle of low self-efficacy and low intrinsic motivation in creative expression.

\subsection{Preliminary Conclusions and Design Implications}
Our study reveals a paradox that prompt-output AI music generation tools boost machine efficiency but undermine adolescent growth, leaving adolescents dependent on outputs and limiting their expressive strategies.

These findings highlight key implications for designing AI music generation tools in adolescent creative education. Tools should scaffold the cognitive shift from vague ideas to structured expressions by making explicit how intentions map onto musical components; they should provide behavioral guidance through iterative refinement that encourages exploration and reflection rather than mere regeneration; and they must balance assistance with autonomy, supporting affective growth by fostering uniqueness, and intrinsic motivation instead of reinforcing dependence on templates.

\section{MusicScaffold}
We propose the MusicScaffold framework, a conceptual workflow that employs explicit guidance and collaborative interaction to support adolescents (ages 12–14) in transforming abstract ideas into structured expressions, thereby cultivating both awareness and strategies of expression.

\subsection{Framework Overview}
\begin{figure*}[htbp]
    \centering
    \includegraphics[width=\linewidth,trim=0 0 0 0,clip]{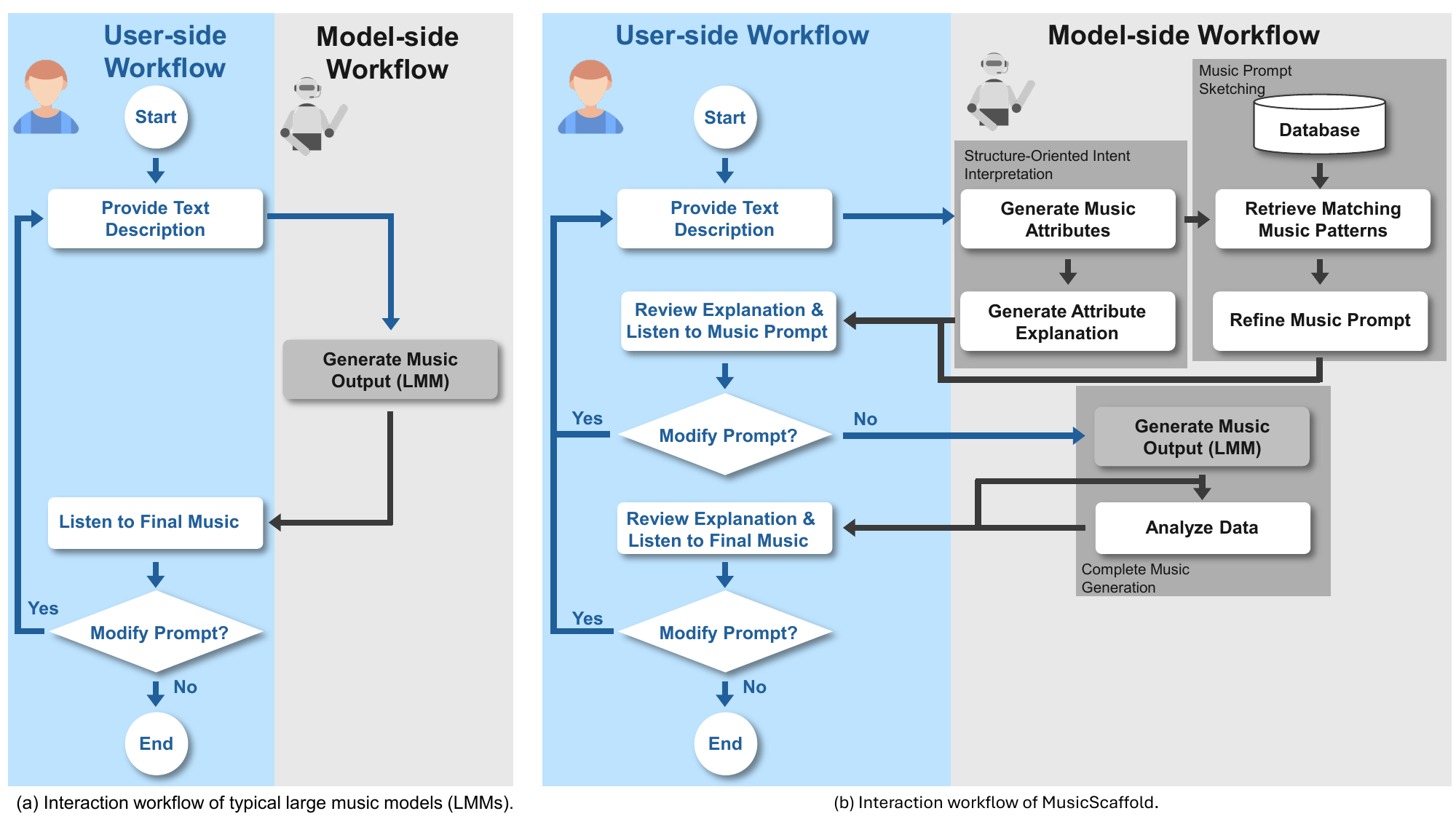}
    \Description{(a) Typical interaction workflow of large music models (LMMs), where a one-shot prompt directly yields an output. (b) The MusicScaffold workflow, which adds symbolic explanation, sketch-based iteration, and reflective feedback to support learning.}
    \caption{Interaction workflows for AI music generation. 
    (a) Typical large music models (LMMs) adopt a direct prompt-output paradigm. 
    (b) MusicScaffold repositions this process as a closed-loop interaction: intent interpretation, music prompt sketching (MPS), and complete music generation, with opportunities to revisit earlier steps and engage in an explain–iterate–co-create loop.}
    \label{fig:method1}
\end{figure*}

Typical large music models (LMMs) follow a one-shot prompt-output paradigm, offering little opportunity for explanation or refinement. In contrast, MusicScaffold integrates user-side and model-side workflows into a closed-loop interaction (Fig.~\ref{fig:method1}). On the user side, learners start with a natural-language description of intent (e.g., ``generate an exciting song''), while on the model side, the input is processed through three stages.

\emph{Structure-oriented intent interpretation.} An LLM interprets vague descriptions into symbolic prompt attributes such as tempo, mode, or rhythmic density. Each attribute is paired with a short explanation of how it connects to the expressed intent (e.g., ``Key: C major. Why? C major sounds bright and simple, which helps the song feel cheerful.''). 

\emph{Music prompt sketching (MPS).} From the interpreted attributes, the system generates symbolic prompts (e.g., MIDI-like representations). These sketches are lightweight drafts that emphasize structural clarity rather than polished audio.

\emph{Complete music generation.} Once the learner confirms a structural plan, the framework invokes a LMM to render a full audio realization. The output is accompanied by reflective feedback, linking the final sound back to the earlier symbolic prompt. 

At any point, learners may return to earlier stages—revising their natural-language intent, reinterpreting attributes, or adjusting sketches. This explain–iterate–co-create loop reflects principles of scaffolding and self-regulated learning that learners externalize reasoning, test adjustments, and progressively internalize expressive strategies. MusicScaffold thus transforms the opaque prompt-output paradigm into an interactive learning environment that supports both efficient creation and long-term growth.

\subsection{Prototype Implementation}
\begin{figure*}[htbp]
    \centering
    \includegraphics[width=\linewidth,trim=0 0 0 0,clip]{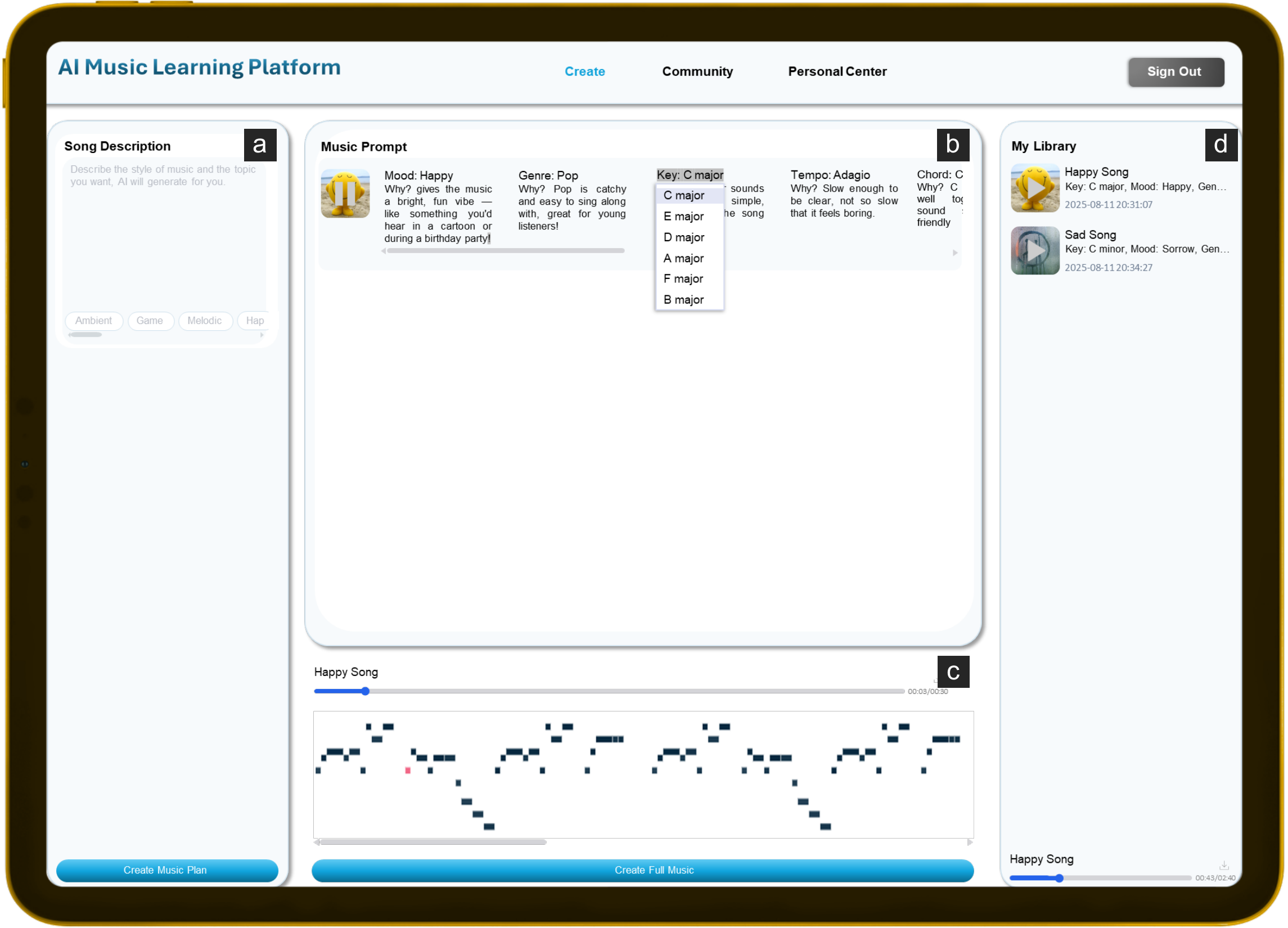}
    \Description{This figure shows the prototype interface of MusicScaffold. 
            It contains four main panels. 
            (a) A text input panel on the left, where students can type free-form natural-language descriptions of their musical intentions or choose from pre-defined templates. 
            (b) An attribute view in the center, which automatically converts the text input into editable musical attributes such as key, tempo, or timbre, each accompanied by a plain-language explanation of its expressive effect. 
            (c) A symbolic prompt preview panel below, which displays these attributes as a short, playable MIDI sketch so students can listen to and iteratively refine their choices. 
            (d) A music library on the right, which archives generated results together with their underlying attribute plans, enabling review, comparison, and reflection over time.}
            
    \caption{Prototype interface of MusicScaffold. (a) Text input panel, where students enter free-form descriptions or select from templates. (b) Attribute view, which translates requests into editable musical attributes with explanations. (c) Symbolic prompt preview, which visualizes attributes as a short, playable MIDI sketch for evaluation and iteration. (d) Music library, which stores generated results together with their corresponding plans for reflection. }
    \label{fig:interface}
\end{figure*}

To examine how MusicScaffold can be operationalized in practice, we built an illustrative prototype of an AI-assisted music learning environment. The goal of this prototype is to instantiate the framework’s core mechanisms in a form accessible to adolescent learners. As shown in Fig.~\ref{fig:interface}, the prototype organizes interaction into a create workspace. This workspace integrates four components: (a) a text input panel for free-form descriptions or starter templates; (b) an attribute view that translates descriptions into symbolic prompt attributes with explanations; (c) a symbolic prompt preview that produces short, editable MIDI sketches; and (d) a music library that archives all generated results with their associated plans, supporting later review and reflection.

\paragraph{Guide.} The system acts as a guide by providing symbolic explanations. When a learner types a vague description such as ``make it sound exciting'', the attribute view (b) decomposes this input into concrete musical elements—such as mood, genre, key, and tempo—each paired with a short natural-language rationale (e.g., ``C major sounds bright and simple, which helps the song feel cheerful''). Alongside each attribute, the system generates a short, playable sketch in (c), so learners can immediately hear how the attribute choice affects musical expression. By explicitly linking affective words to structural features, the system guides students from raw impulses to structured design elements, giving them a clear path forward.

\paragraph{Coach.} The system also acts as a coach by prompting reflection during iterative refinement. After receiving initial explanations, learners can directly adjust attributes in the attribute view (b). Each adjustment generates a new short sketch and appends a reflective question to the modified attribute (e.g., ``Does the slower tempo still convey the brightness you intended?’’ when tempo is changed). These questions encourage learners to articulate why certain structures fit or fail, turning each modification into an opportunity for reasoning. Through this cycle of trial, reflection, and adjustment, the system coaches learners to move beyond one-shot acceptance and to build deliberate expressive strategies.

\paragraph{Partner.} Finally, the system acts as a partner by supporting co-creation while safeguarding learner autonomy. Once the learner confirms a structural plan, the framework invokes the LMM to render a full audio piece. Crucially, the system does not overwrite the learner’s choices but the generated output is always tied back to the student’s selected attributes and prior sketches. For instance, if a student chose a minor mode to express ``sad but hopeful’’, the final piece is produced in line with this decision and returned with an explanatory report linking the outcome to the student’s input. The learner can then decide to accept, reject, or further refine the piece. In this way, the system co-creates music that reflects the student’s intentions, while preserving agency—ensuring that adolescents remain the primary creators rather than passive consumers of AI outputs.

\subsection{Technical Implementation}

\begin{figure*}[htbp]
    \centering
    \includegraphics[width=\linewidth,trim=0 0 0 0,clip]{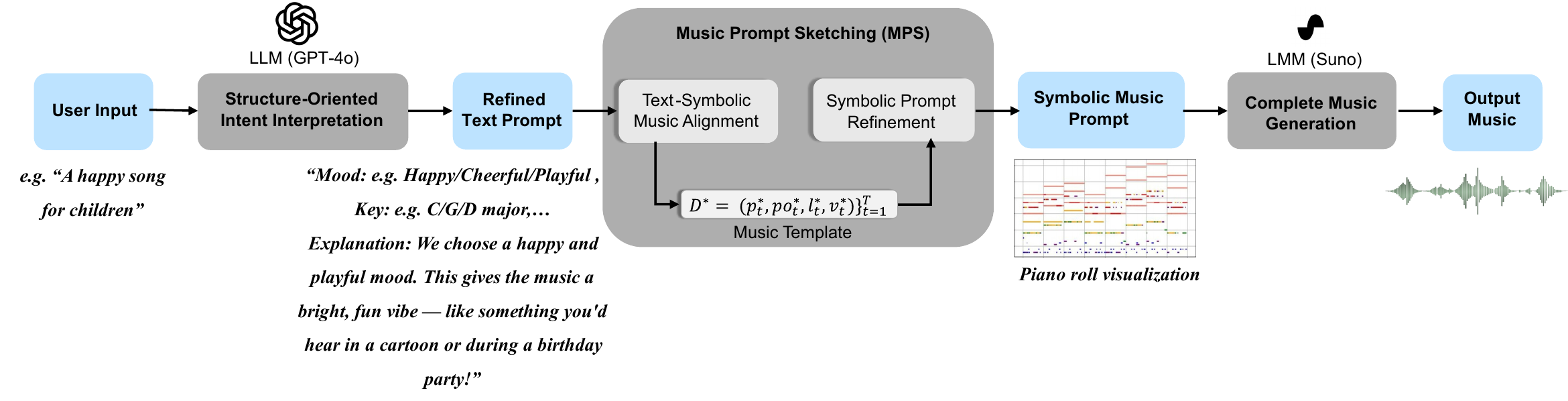}
    \Description{This figure illustrates the technical implementation pipeline of MusicScaffold. 
It is organized into three sequential stages. 
First, structure-oriented intent interpretation: natural-language input from the learner is processed by a large language model to extract symbolic attributes such as mood, key, or tempo, each paired with plain-language explanations. 
Second, music prompt sketching: the extracted attributes are aligned with candidate symbolic segments from a database, then refined using music-theory rules to form an editable symbolic prompt represented as a short MIDI sketch. 
Third, complete music generation: the refined symbolic prompt is sent to a large music model to produce high-fidelity audio. 
The resulting output is stored in the music library together with an explanatory alignment report. 
Arrows between stages indicate that learners can loop back to earlier steps, making each computational process transparent, editable, and supportive of iterative learning.}

    \caption{Technical implementation of MusicScaffold. The pipeline connects three stages: (1) structure-oriented intent interpretation, where natural-language inputs are mapped to symbolic attributes with explanations; (2) music prompt sketching, where attributes are aligned and refined into editable symbolic prompts; and (3) complete music generation, where the refined prompt is rendered by a large music model (LMM) and stored with an explanatory report in the music library. This design makes each computational step visible and editable, supporting transparency and iterative learning.}
    \label{fig:method3}
\end{figure*}

The technical design of MusicScaffold mirrors the three layers introduced in the framework overview and is instantiated through the prototype interface (Fig.~\ref{fig:method3}). Each computational step is surfaced in the interface to ensure transparency, editability, and iterative learning support.

\emph{Structure-oriented intent interpretation.} Given a natural-language input \(u \in \mathcal{U}\), the LLM produces a structured attribute set

\begin{equation}
v = M_{\text{LLM}}(u) = \{(x_i, \alpha_i)\}_{i=1}^K ,    
\end{equation}
where each attribute \(x_i\) (e.g., mood, key, tempo) has a value \(\alpha_i\). We classify them as descriptive \(\mathcal{X}_d\) (mood, genre), global \(\mathcal{X}_g\) (tempo, key), and local \(\mathcal{X}_l\) (chord sequences, rhythms). These attributes are surfaced in the attribute view with explanations, making the model’s interpretation transparent for adolescents.

\emph{Music prompt sketching (MPS).} This stage transforms attributes into an editable symbolic prompt through two modules:
\begin{itemize}
    \item \emph{Text–symbolic alignment:} A lightweight retrieval model aligns the attribute set \(v\) with a candidate symbolic segment in a database \(\mathcal{D}\). Formally:
    \begin{equation}
    D^* = \arg\max_{D_i \in \mathcal{D}} \mathrm{R}(v, D_i), \quad
    \mathrm{R}(v, D_i) = \sum_{(x_j, \alpha_j) \in v} w_j\, \mathbb{I}(D_i[x_j] = \alpha_j),
    \end{equation}
    where \( w_j \in [0,1] \) is the attribute weight, \(\mathbb{I}(\cdot)\) is an indicator function and \(D^*\) is the best-matching segment.
    \item \emph{Refinement:} The retrieved segment is adapted using music-theory rules \(r\) inferred by the LLM:
    \begin{equation}
    s = \Gamma(D^*, v, r),
    \end{equation}
    with \(D^*\) represented as note tuples \( \{(p_t^*, po_t^*, l_t^*, v_t^*)\}_{t=1}^{T}\) (Pitch, Position, Length, Velocity) where $t$ is the bar index. The refinement step adjusts harmony, rhythm, or melodic contours to align more closely with the intended attributes.
\end{itemize}

The resulting symbolic prompt \(s\) is rendered as a playable MIDI excerpt in the prompt panel, allowing learners to explore, modify, and understand the link between attributes and musical form before committing to full audio generation.

\emph{Complete music generation.} Once the sketch is confirmed, it is passed to a large music model (LMM) for rendering:
\begin{equation}
o = M_{\text{LMM}}(s),
\end{equation}
producing final audio \(o \in \mathcal{O}\). The system then checks alignment between \(o\) and \(v\), generating an explanatory report stored alongside the audio in the music library. If misalignment or dissatisfaction occurs, the process loops back, sustaining the explain–iterate–co-create cycle.

\emph{Process summary.} The full pipeline is expressed as:
\begin{equation}
u \xrightarrow{M_{\text{LLM}}} v \xrightarrow{\text{MPS}} s \xrightarrow{M_{\text{LMM}}} o .
\end{equation}

This mapping connects technical modules to interface components, ensuring that the computational backbone supports educational goals of transparency, editability, and iterative creative learning.

\section{Comparative Study of Generative AI Paradigms in Adolescent Creative Education}

\subsection{Study Design}
\subsubsection{Participants and Context}
A total of 270 middle school students ($N=270$, ages 12–14) participated, drawn from nine parallel music classes in the same school. Each class was randomly assigned to one of the three experimental conditions, resulting in 90 students per group. All students had comparable exposure to basic music education (general music appreciation courses) but no formal training in composition. Prior to the study, the teacher confirmed through a short baseline survey that there were no significant differences in students’ prior musical background or creative self-efficacy across groups.
 
The study was embedded in regular music classes, following the same four-week duration, weekly 30-minute sessions and after-class activity as in the exploratory study. This ensured comparability while allowing systematic contrasts across conditions.

\subsubsection{Procedure}
The weekly procedure followed the same three stages as in the exploratory study—task introduction, student creation, and feedback collection—plus the optional after-class activity. The key difference across conditions lay in the tools assigned to each group:

\begin{itemize}[left=0pt]
        \item Control Group A (Direct Generation): Students used existing commercial AI music products. These tools represent the typical prompt-output paradigm: users input a free-form text description, and the system generates a complete song in a single pass, without symbolic prompt explanation or iterative refinement.  
        \item Control Group B (Attribute Selection): Students used the same commercial products, but interacted via fixed attribute sliders (e.g., tempo, style, timbre). This condition reflects the parameterized editing paradigm, offering structured but pre-defined options rather than free text.  
     \item Experimental Group C (MusicScaffold Framework): Students used our prototype implementing explain–iterate co-creation, where free-form inputs were mapped to symbolic musical attributes with plain-language explanations. Students could iteratively refine these attributes and preview symbolic prompts before committing to full generation.  
\end{itemize}
At the end of the final week, teachers also conducted short semi-structured interviews with 12 randomly selected students (4 per group). Questions asked students to reflect on what they learned, how the tools supported or limited their expression, and whether they felt more capable of shaping music intentionally. These conversations provided supplementary qualitative data contextualizing survey and log analyses. To further ensure validity, we invited education experts to review the questionnaire for conceptual clarity and domain relevance, and survey methodology experts to assess its scientific soundness and feasibility. We manually checked all responses and found no missing data, resulting in 270 valid questionnaires.

\subsubsection{Analysis Approach}
We used a mixed-methods approach. Quantitatively, categorical and ordinal data (e.g., specificity, coverage, behaviors, originality, self-efficacy) were analyzed using chi-square and Kruskal--Wallis tests, with Dunn’s or Mann--Whitney U post-hoc comparisons and Bonferroni correction. Longitudinal measures (e.g., after-class interactions) were examined week by week with Kruskal--Wallis tests and within-group trend analyses. Qualitatively, open-ended survey responses were thematically coded to capture students’ perceptions of strategy, autonomy, and tool support, providing context for the quantitative results.

\subsection{Results}
\begin{figure*}[htbp]
    \centering
    \includegraphics[width=\linewidth,trim=0 0 0 0,clip]{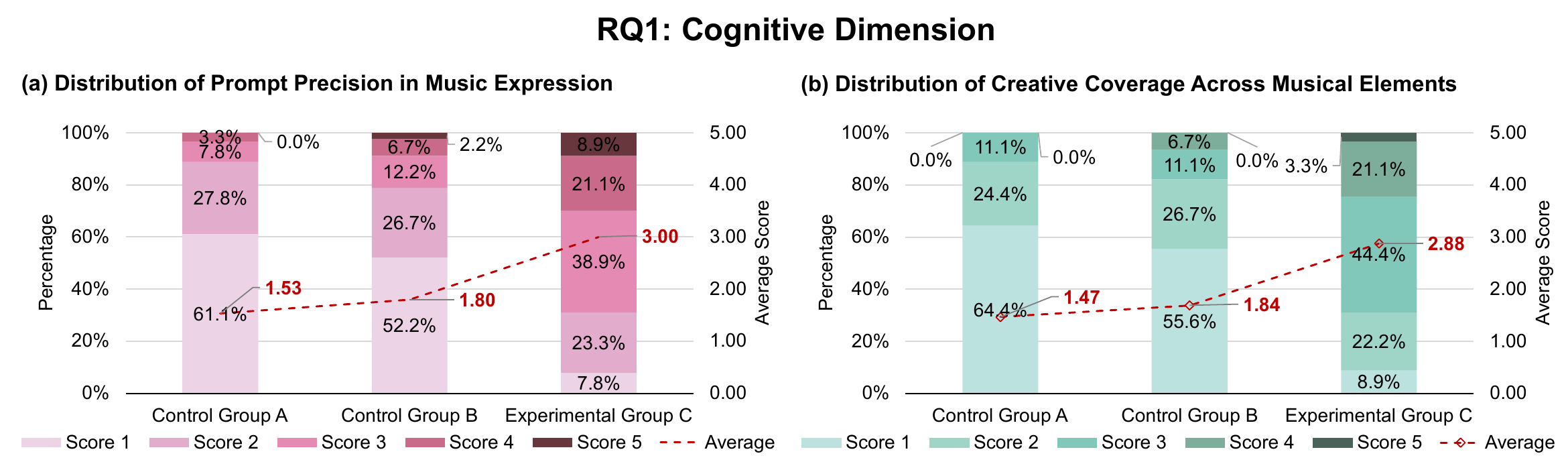}
    \Description{Results for RQ1: Prompt specificity and element coverage across three groups. MusicScaffold students reached higher levels of specificity and invoked broader musical dimensions compared to both control groups.}
    \caption{RQ1: MusicScaffold significantly improves adolescents’ recognition and expressive abilities. 
    (a) Distribution of prompt specificity scores, showing that Group~C (MusicScaffold) produced more Level~3–5 descriptions compared to Groups~A and~B. 
    (b) Breadth of musical elements used, where Group~C engaged with multiple dimensions (e.g., rhythm, mode, timbre) far more often than controls.}
    \label{fig:main1}
\end{figure*}
\subsubsection{RQ1: MusicScaffold significantly improves adolescents’ recognition and expressive abilities}

Fig.~\ref{fig:main1} (a) presents the distribution of prompt specificity scores. In Control Group~A (prompt-output), 88.9\% of inputs were rated at Level~1 and Level~2. Control Group~B (attribute-selection) showed a slightly higher distribution, with 78.9\% at Level~1 and Level~2. In contrast, Experimental Group~C (MusicScaffold framework) exhibited a striking shift: only 31.1\% of inputs were Level~1 and Level~2, while 38.9\% reached Level~3, 21.1\% Level~4, and 8.9\% Level~5. The average score for Group~C (3.00) was nearly double that of Group~A (1.53) and Group~B (1.80).  

Similarly, Fig.~\ref{fig:main1} (b) shows the breadth of musical elements used. Control~A’s prompts overwhelmingly focused on one dimension (64.4\%), with an average score of 1.47. Control~B displayed a marginal increase (mean~=~1.84), but still remained narrow in scope. In contrast, Group~C achieved significantly broader coverage: 68.8\% of prompts invoked three or more dimensions, with an average score of 2.88.  

Statistical analyses confirmed these differences. A chi-square test revealed significant group effects for both prompt specificity ($\chi^2(8, N=270)=91.50$, $p<.001$) and element coverage ($\chi^2(8, N=270)=96.31$, $p<.001$). Nonparametric rank-based comparisons (Kruskal--Wallis) corroborated the findings (specificity: $H(2)=65.71$, $p<.001$; coverage: $H(2)=91.50$, $p<.001$). Pairwise comparisons (Mann--Whitney U with Bonferroni correction) showed no significant difference between Control~A and Control~B ($p_{adj}>.3$), but highly significant differences between both control groups and the Experimental Group~C ($p_{adj}<.001$).  

Beyond the statistics, students’ subjective reflections provide further insight. In Control~A and B, many students expressed frustration: \textit{``I just kept writing `happy' or `sad', but the music never really sounded right''} (Control~A), or \textit{``I clicked the attributes, but I didn’t really know what each meant, so I just picked something random''} (Control~B). By contrast, Group~C students reported a growing awareness of strategy: \textit{``I realized if I add staccato it makes the rhythm more lively, and then I can change the key to make it sound brighter''}. Another noted: \textit{``At first I only thought about speed, but after trying different chord progressions and rhythmic patterns, I found more ways to make the music fit my idea''}.  

Taken together, these results demonstrate that adolescents in both control conditions largely remained at a vague, single--dimensional level of expression. In contrast, the MusicScaffold framework significantly enhanced their ability to translate creative intentions into structured, multi--dimensional strategies, moving them beyond intuitive labels toward deliberate expressive design.

\subsubsection{RQ2: MusicScaffold fosters strategic adjustments and reduces homogeneity over time}
\begin{figure*}[htbp]
    \centering
    \includegraphics[width=\linewidth,trim=0 0 0 0,clip]{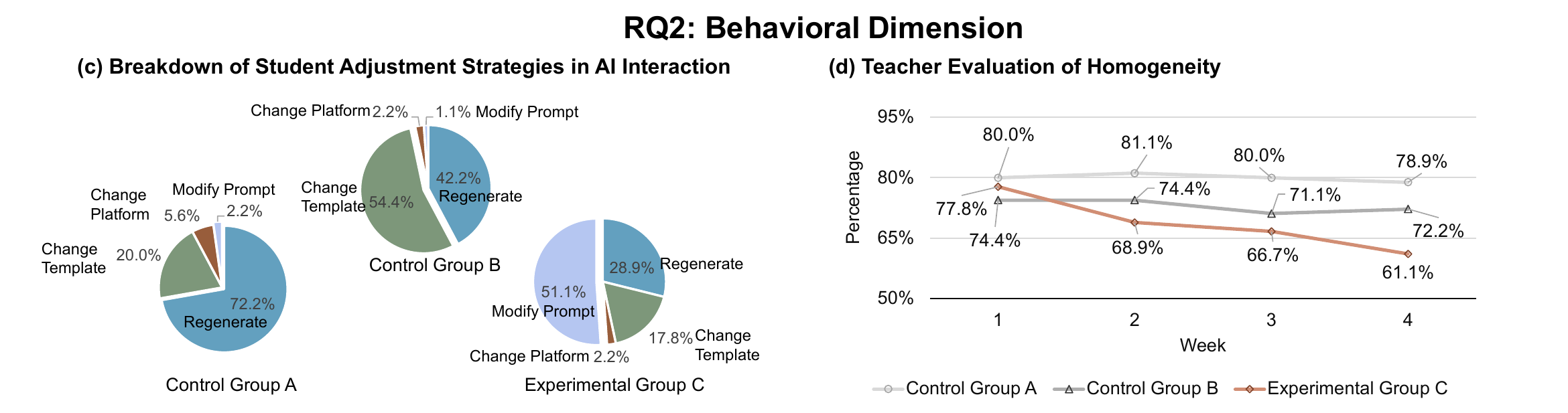}
    \Description{Results for RQ2: Adjustment behaviors and prompt homogeneity across four weeks. MusicScaffold encouraged more strategic refinements and reduced homogeneous outputs over time.}
    \caption{RQ2: MusicScaffold fosters strategic adjustments and reduces homogeneity over time. 
    (a) Adjustment behavior distribution, with Group~C showing a clear shift toward prompt modifications instead of repeated regenerations. 
    (b) Teacher evaluations of prompt homogeneity, where only Group~C demonstrated reduced reliance on repetitive input–output mappings by Week~4.}
    \label{fig:main2}
\end{figure*}
Fig.~\ref{fig:main2} (a) shows students’ adjustment behaviors. Control Group~A relied heavily on non-strategic regeneration (76.7\%), with only 2.2\% prompt refinements. Control~B showed a similar pattern (42.2\% regenerations, 54.4\% template switches, 1.1\% refinements). In contrast, Experimental Group~C shifted markedly, with 51.1\% prompt modifications and only 28.9\% regenerations. Group differences were significant ($\chi^2(6, N=540)=194.27$, $p<.001$); post-hoc tests showed no difference between A and B, but both differed strongly from C ($p_{adj}<.001$). Thus, while baseline systems reinforced trial-and-error, MusicScaffold promoted reflective and strategic refinement.  

Fig.~\ref{fig:main2} (b) presents teacher evaluations of the homogeneity of students’ behavior across four weeks. In Week~1, no significant group differences were observed ($\chi^2(2, N=270)=0.80$, $p=.67$), confirming that all groups started from a comparable baseline. By Week~2 and Week~3, overall chi-square tests showed emerging but non-significant trends (Week~2: $p=.17$; Week~3: $p=.12$). Crucially, by Week~4, a significant group effect emerged ($\chi^2(2, N=270)=7.01$, $p=.03$). Post-hoc comparisons revealed that Experimental Group~C differed significantly from Control~A ($p_{adj}=.044$), while A and B remained indistinguishable ($p_{adj}>.3$). This indicates that only the MusicScaffold framework succeeded in reducing over-reliance on homogeneous input–output mappings across time.  

Students’ subjective reflections further corroborate these behavioral patterns. In Control Group~A, one participant noted: \textit{``If I didn’t like it, I just kept clicking regenerate until something sounded better''}. A Control~B student admitted: \textit{``I changed the style tags a few times, but I wasn’t sure what they really meant''}. In contrast, students in Group~C increasingly described purposeful strategies: \textit{``When I wanted more tension, I tried adding a minor chord and slowing the tempo''}; \textit{``I learned that changing timbre, not just speed, could make the mood clearer''}. These reflections reveal not only more diverse behaviors but also a growing meta-awareness of expressive mechanisms.

The behavioral and evaluative evidence highlights that while conventional prompt-output and attribute-selection systems entrenched non-strategic habits, the MusicScaffold framework enabled adolescents to shift toward reflective, strategy-based adjustments. Over multiple weeks, this led to a measurable reduction in homogeneous outcomes, suggesting that MusicScaffold can foster sustainable creative growth rather than superficial reliance on automated outputs.

\subsubsection{RQ3: MusicScaffold strengthens adolescents’ autonomy, confidence, and sustained motivation in creative expression}

\begin{figure*}[htbp]
    \centering
    \includegraphics[width=\linewidth,trim=0 0 0 0,clip]{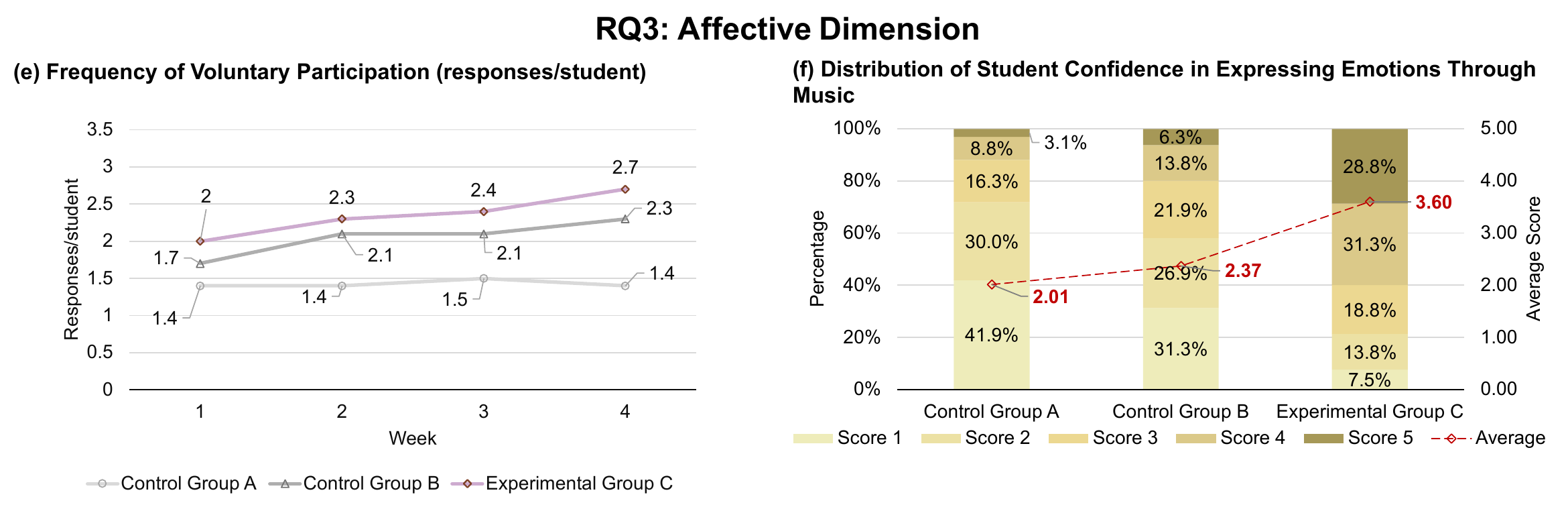}
    \Description{Results for RQ3: Out-of-class participation and self-efficacy. MusicScaffold increased voluntary engagement and strengthened students’ confidence in musical expression.}
    \caption{RQ3: MusicScaffold strengthens adolescents’ autonomy, confidence, and sustained motivation. 
    (a) After-class participation, where Group~C consistently outperformed controls across four weeks. 
    (b) Self-efficacy ratings, with Group~C reporting higher confidence in expressing emotions through music than Groups~A and~B.}
    \label{fig:main3}
\end{figure*}

Fig.~\ref{fig:main3}(a) shows students’ out-of-class participation. In Week~1, all groups started at a low mean value (A: $M=1.4$, B: $M=1.7$, C: $M=2$), with no significant differences except A vs.\ C ($p_{adj}=.0015$). From Week~2 onward, Group~C rose sharply ($M=2.7$) and consistently outperformed both controls ($p_{adj}<.01$), while A and B remained flat. Trend analyses confirmed no growth for Group~A ($p=.81$), modest gains for Group~B ($p=.008$), and a strong upward trajectory for Group~C ($p=.017$), whose participation more than doubled from Week~1 to Week~4.

Fig.~\ref{fig:main3}(b) shows students’ confidence in musical expression. Control~A stayed low (71.9\% at Scores 1–2, $M=2.01$), Control~B improved moderately ($M=2.37$), while Group~C reached the highest self-efficacy ($M=3.60$), with more students at Levels 3–5. Statistical tests confirmed significant group effects ($\chi^2(8, N=270)=33.89, p<.001$; $H(2)=27.97, p<.001$). Pairwise comparisons showed both B and C significantly outperformed A ($p_{adj}<.001$), but did not differ from each other ($p_{adj}=.93$).

Beyond statistics, students’ subjective reflections highlight the qualitative difference. In Control~A, many expressed discouragement: \textit{``Even if I use the tool, the music doesn’t feel like what I really want, so I don’t bother trying more''}. Control~B students reported slightly higher confidence, but often tied it to the system’s default structures: \textit{``I clicked some attributes and it felt better, but I wasn’t sure why''}. By contrast, Group~C students described stronger motivation: \textit{``I wanted to guess how others used staccato or chords to show emotions, and then try it myself''}, and \textit{``After practicing with MusicScaffold, I feel I can use rhythm and harmony to make my idea clearer''}.  

These results suggest that while standard prompt-output AI music generation tools (Group~A) and attribute-selection tools (Group~B) provide limited support for adolescents’ creative autonomy, the MusicScaffold framework significantly enhances both sustained motivation and self-efficacy. Through guided symbolic prompt explanation and iterative refinement, students not only increased their willingness to participate in creative activities, but also developed greater confidence in their ability to express emotions through music.

\subsection{Case Study: Student L’s Growth with MusicScaffold}
Student L (age 13), a second-year middle school student, began by typing ``happy song'' in the first class. The first turning point came when the system returned happy (commonly linked to major key, fast tempo, bright timbre) with short explanations for each element. Curious, L experimented by changing fast tempo to medium, but after hearing the slower result switched back, commenting that ``fast really makes it feel lively''. This showed how symbolic prompt explanations immediately nudged L from vague emotion to structural awareness. By Week 3, L faced difficulty expressing sadness. Using the history view in the Music Library, L compared earlier attempts and noticed the consistent link: minor key → melancholic tone. Acting on this explanation, L revised the prompt to ``sad feeling in minor key, with slow tempo and soft piano''. This marked the first time L deliberately coordinated multiple attributes, connecting system-provided explanations with strategic refinement. In Week 4, after several iterations guided by iterative refinement prompts (e.g., highlighting swing rhythm vs. straight rhythm as options), L confidently articulated: ``A jazz ballad with melancholic; a minor key; slow tempo with a swing rhythm''.

Behaviorally, L’s early reliance on blind regenerations gave way to targeted refinements once the Music Library enabled structured comparison. Affective growth was tied to specific interactions: after failing three times with ``sad'' in major key, L finally succeeded by adjusting to minor key with system feedback, exclaiming ``Now I know what to fix, not just guess''. These shifts illustrate how MusicScaffold’s symbolic prompt explanations and iterative refinement mechanisms directly fostered cognitive, behavioral, and affective growth, enabling learners to progress from vague intent to structured expression, from trial-and-error to strategic regulation, and from frustration to confident agency.

\section{Discussion}
\subsection{Extending Educational Theory in the Era of Generative AI}
Our findings contribute to broader debates in educational psychology and creativity research by showing how generative AI reshapes long-standing theories. The results demonstrate that generative AI can be more than a content generator. When designed with multi-faceted roles, it actively mediates adolescents’ transition from intuitive inspiration to deliberate strategy. This resonates with \emph{scaffolding} theory \cite{vygotsky1978mind,bruner1974toward}, yet also extends it. Whereas traditional scaffolding presumes guidance from a human teacher or expert, MusicScaffold illustrates how hybrid AI–human scaffolds can externalize reasoning in ways that are explicit, repeatable, and accessible even in teacher-limited contexts. This suggests a new locus of scaffolding, where responsibility for making tacit strategies visible is shared between humans and machines.

The framework also enriches \emph{self-regulated learning} theory \cite{zimmerman2002becoming}. Classic accounts emphasize learners’ own planning, monitoring, and reflection, but MusicScaffold positions AI as a metacognitive artifact that prompts and sustains these processes. By providing symbolic prompt explanations and iterative refinement, the system transforms this theory from an exclusively individual practice into a form of ``co-regulated learning'', in which learners and AI jointly manage the cycle of strategy use and reflection. This reconceptualization highlights AI’s potential to scaffold not only content knowledge but also the development of regulatory habits that underpin long-term growth.

Finally, the observed gains in confidence and voluntary engagement align with \emph{self-determination theory} \cite{ryan2000self}, but point toward a novel design stance we term \emph{structured autonomy}. Rather than treating autonomy as the absence of external control, MusicScaffold shows how AI can provide structured pathways without undermining learners’ agency. By guiding expressive choices, the system fosters competence and motivation simultaneously, offering a distinctive model of autonomy-supportive AI design. 

Taken together, these insights suggest that generative AI does not simply instantiate existing theories in a new domain, but also challenges us to reconsider foundational concepts: the source of scaffolding, the nature of self-regulation, and the conditions under which autonomy is preserved. In this sense, MusicScaffold extends classic educational theories into the era of generative AI, while also opening new conceptual ground for how human–AI systems can support adolescent creative development.

Beyond theory, our study points to concrete directions for designing generative AI systems in education. First, systems should incorporate symbolic prompt explanation modules, turning abstract prompts into interpretable components (e.g., rhythm, harmony, mood) that learners can manipulate directly. Second, interfaces should embed reflective feedback mechanisms, such as history tracking or side-by-side comparisons, that nudge students toward deliberate refinements rather than repeated regenerations. Third, future systems should offer adjustable guidance levels, enabling adolescents to gradually assume more control as their confidence grows, thereby balancing system support with learner autonomy. Taken together, these design principles highlight how generative AI can evolve from passive output tools into active learning supports that cultivate sustainable creative growth.

\subsection{Human Strengths and Machine Trade-offs in Adolescent Creative Support}
A central question in the design of educational generative AI is how its role compares with that of human educators. While both humans and machines can facilitate creative expression, their mechanisms and strengths differ in important ways.

Human educators bring unique capabilities that AI cannot replicate \cite{harvey2025don,wragg2025investigating,sun2025live}. Teachers and peers are able to perceive learners’ nuanced needs, frustrations, and latent intentions through multimodal cues such as facial expression, body language, and social context. They also provide humanistic care and social connection, which foster belonging and emotional safety—critical ingredients for sustained creative exploration. Moreover, human guidance embeds cultural meaning and values into learning processes, helping adolescents situate creative expression within broader personal and societal narratives.

By contrast, AI systems offer distinctive advantages in lowering barriers for novices. Generative tools reduce the pressure of early failure by providing immediate, polished outputs, which can help beginners build confidence and maintain motivation. They create a low-stakes environment where adolescents can experiment without fear of judgment, and they expose students to complete creative products earlier than traditional instruction would allow. Such immediacy can be especially valuable for sustaining engagement in the initial stages of skill acquisition.

The educational challenge, therefore, lies in balancing these complementary roles. AI should be positioned as a supportive amplifier—providing accessible, low-pressure entry points and rapid feedback—while human educators remain essential for observing learners’ deeper needs, nurturing autonomy, and guiding long-term ability growth. Rather than replacing teachers, AI can complement them by filling gaps in scale and immediacy, while humans ensure that creative learning remains culturally grounded, emotionally supportive, and developmentally meaningful.

\subsection{Limitations and Future Work}
While our findings highlight the potential of the MusicScaffold framework, several limitations warrant consideration. First, our study was conducted with adolescents in one middle school, which may limit generalizability across broader cultural and educational contexts; future work should test the framework in diverse age groups and learning settings. Second, although we measured changes across four weeks, longer-term studies are needed to understand whether the observed gains in strategy use, self-regulation, and motivation are sustained over time. Third, our evaluation primarily focused on music, a domain with relatively clear mappings between emotions and structure; applying MusicScaffold to domains such as visual art or writing may reveal new challenges where symbolic prompt mappings are less precise. Finally, while we captured both behavioral logs and self-reports, more fine-grained process data (e.g., think-aloud protocols, longitudinal interviews) could deepen our understanding of how adolescents internalize explanatory guidance. Addressing these limitations will help clarify the boundary conditions of MusicScaffold and refine its applicability as a general model for AI-supported creative learning.

\subsection{Ethical Considerations}
This study was conducted in collaboration with the partner middle school and approved by its ethics and review board in accordance with local educational regulations. Participation was voluntary, with informed consent obtained from both students and their guardians. Teachers emphasized that involvement was unrelated to students’ grades or classroom standing, and students could withdraw at any time without consequence. To protect privacy, no personal identifiers were collected; all prompts, outputs, and survey responses were anonymized before analysis and stored on secure school servers. For music generation, we used Suno \cite{suno2024} and Udio \cite{udio}, following their publicly documented interfaces. These tools were used strictly for research purposes within the classroom study, and all generated music as well as interaction logs were retained only for analysis in this project. No content was shared beyond the study context, and no commercial use was involved. We followed the platforms’ terms of service and ensured that the integration preserved both lawful use and the integrity of participants’ data. 


\section{Conclusion}
MusicScaffold demonstrates that generative AI can be repositioned from an output-oriented tool to a learning companion that cultivates adolescents’ expressive growth. Rather than emphasizing efficiency alone, the framework highlights how symbolic explanations and iterative refinement can support the transition from vague impulses to structured creativity. Our studies show that this design strengthens adolescents’ cognitive specificity, behavioral regulation, and affective confidence, suggesting that scaffolding-based AI can foster both competence and motivation. 

At a broader level, these findings extend theories of scaffolding and self-regulated learning into the context of generative AI and suggest a design perspective in which technology supports human development rather than replacing it. Future work should examine how this paradigm can be applied to other creative domains, how it operates over longer periods in classrooms, and how teachers and peers can collaborate with AI to shape supportive learning environments.



\bibliographystyle{ACM-Reference-Format}
\bibliography{sample-base}

\appendix

\end{document}